\documentclass{ws-procs9x6}

\usepackage{amsmath, amssymb}
\usepackage{graphicx}
\newcommand{\vx}{\mathbf{x}}
\newcommand{\vy}{\mathbf{y}}
\newcommand{\vz}{\mathbf{z}}
\newcommand{\vu}{\mathbf{u}}
\newcommand{\vv}{\mathbf{v}}

\newcommand{\abar}{\bar\alpha}
\newcommand{\avg}[1]{\left\langle #1 \right\rangle}
\newcommand{\erfc}{\text{erfc}}

\begin{document}

\title{From QCD at high energy to statistical physics and back}

\author{G. Soyez\footnote{On leave from the fundamental theoretical physics group of the University of Li\`ege.}}

\address{SPhT, CEA/Saclay, Orme des Merisiers, B\^at 774, F-91191 Gif-sur-Yvette cedex, France}

\maketitle

\abstracts{
In these proceedings, we shall first recall the evolution equations arising when increasing the rapidity $Y=\log(s)$ within the perturbative QCD regime. We shall then summarise the main properties on their asymptotic solutions and discuss the physical picture emerging from our analysis.}

We start with the equations describing the evolution of scattering amplitudes towards high-energy. For simplicity, we shall work in the large-$N_c$ limit for which we can use the dipole picture. Thus, we consider the scattering amplitude $\avg{T^{(k)}_{\vx_1\vy_1;...;\vx_k\vy_k}}$ between a projectile made of fast-moving colourless $q\bar q$ dipoles of transverse coordinates $(\vx_i, \vy_i)$ and a generic target. 

The evolution equation for $\avg{T_{\vx\vy}}$ can be obtained from dipole splitting in the projectile (the large-$N_c$ version of a gluon emission) which directly leads to
\begin{equation}\label{eq:bk}
\partial_Y \avg{T_{\vx\vy}} = \frac{\abar}{2\pi}\int d^2z\, \frac{(\vx-\vy)^2}{(\vx-\vz)^2(\vz-\vy)^2}
\left[ \avg{T_{\vx\vz}} + \avg{T_{\vz\vy}} - \avg{T_{\vx\vy}} - \avg{T^{(2)}_{\vx\vz;\vz\vy}}\right].
\end{equation}
The three linear terms in this equation correspond to the well-known BFKL equation. This has solutions growing like $\exp(\omega_P Y)$ (with $\omega_P=4\log(2)\abar$) and thus violating the unitarity bound $T_{\vx\vy}\le 1$. The last term, taking into account the scattering on both dipoles, is important when the scattering becomes of order 1 and restores unitarity. 
To deal with \eqref{eq:bk}, one usually adopts in the {\em mean-field approximation} $\avg{T^{(2)}_{\vx\vz;\vz\vy}}=\avg{T_{\vx\vz}}\avg{T_{\vz\vy}}$ which gives a closed equation known as the Balitsky-Kovchegov (BK) equation.

In general, one has to consider an infinite hierarchy of equation for $\avg{T^{(k)}}$. From dipole splitting in the projectile, one gets a BFKL-type contribution proportional to $\avg{T^{(k)}}$ and saturation corrections going like $\avg{T^{(k+1)}}$. The resulting infinite set of equations is known as the (large-$N_c$) Balitsky hierarchy. 
Nevertheless, it has been shown recently that additional contributions corresponding to {\em gluon-number fluctuations} have to be included. Those can be computed through dipole splitting in the target and give an additional contribution proportional to $\avg{T^{(k-1)}}$ to the evolution of $\avg{T^{(k)}}$. For the second equation in the hierarchy, this is
\begin{eqnarray}\label{eq:fluct}
\lefteqn{\left.\partial_Y\avg{T^{(2)}_{\vx_1\vy_1;\vx_2\vy_2}}\right|_{\text{fluct}}
 = \frac{1}{2}\frac{\abar}{2\pi}\left(\frac{\alpha_s}{2\pi}\right)^2}\\
&& \int_{\vu\vv\vz} {\mathcal{M}}_{\vu\vv\vz} {\mathcal A}_0(\vx_1\vy_1|\vu\vz){\mathcal A}_0(\vx_2\vy_2|\vz\vv) \nonumber
   \nabla_\vu^2 \nabla_\vv^2 \avg{T_{\vu\vv}} + (1 \leftrightarrow 2),
\end{eqnarray}
where $\mathcal{M}$ is the dipole splitting kernel and $\mathcal{A}_0$ the dipole-dipole scattering amplitude.
The new term becomes comparable to the BFKL contribution when $T\sim \alpha_s^2$ {\em i.e.} in the dilute regime as expected. Also, by combining pomeron splittings and mergings we can build pomeron loops. Note finally that, even if one has to pay the price of an additional factor $\alpha_s^2$, once we have a pomeron splitting the amplitude grows like two BFKL pomerons. This becomes comparable to the one-pomeron-exchange for $Y\gtrsim \frac{1}{\omega_P}\log(1/\alpha_S^2)$. We shall discuss in more details the consequences of the fluctuation term later on.

Before considering the solutions of those equations, let us quote that the full set of equations can be seen \cite{ist} as a reaction-diffusion process in which one dipole $(\vx,\vy)$ can split into two dipoles $(\vx,\vz)$ and $(\vz,\vy)$ through usual BFKL splitting and two dipoles $(\vx_1,\vy_1)$, $(\vx_2,\vy_2)$ can merge into one dipole $(\vu,\vv)$ through a vertex directly extracted from \eqref{eq:fluct}. However, the merging vertex is not positive definite, which proves that this dipole model is only effective and fluctuations really involve the quantum gluonic degrees of freedom.

Let us now discuss the asymptotic solutions of those equations. We shall neglect impact parameter and work in momentum space $k$. We start with the BK equation which has been shown to lie in the same universality class as the Fisher-Kolmogorov-Petrovsky-Piscounov (F-KPP) equation known since seventy years in statistical physics. This equation admits travelling waves as asymptotic solutions. In terms of QCD variables, this translates into the {\em geometric scaling} property, stating that $T(k,Y)$ is a function of the ration $k/Q_s(Y)$ only, where $Q_s(Y)$ is the {\em saturation scale} which provides a natural infrared regulator. The analogy with the F-KPP equation predicts that $T(k,Y)\sim [k^2/Q_s^2(Y)]^{-\gamma_c}$ for $k\gg Q_s$, with $Q_s^2(Y)=k_0^2\exp(\abar v_c Y)$. $\gamma_c$ and $v_c$ are pure numbers determined from the BFKL kernel only. It is important to notice that the property of geometric scaling is a consequence of saturation which extends (as $\sqrt{Y}$) far above the saturation domain, where $T\ll 1$.

If we switch on the fluctuation term, the situation becomes a bit more intricate. However, neglecting the impact-parameter dependence, one can perform a local-noise approximation\footnote{A study of the impact-parameter averaged equations beyond the local-noise approximation is under study using a properly defined matching between a mean-field BK evolution close to the unitarity limit and random dipole splitting in the dilute tail \cite{imms}.} which allows to rewrite the infinite hierarchy for average amplitudes as a single Langevin equation for event-by-event amplitudes which is, formally, the BK equation supplemented by a noise term
\begin{equation}\label{eq:langevin}
\partial_Y T(k,Y) = \abar K_{\text{BFKL}} \otimes T(k,Y) -\abar T^2(k,Y) + \abar \sqrt{2\kappa\alpha_s^2T(k,Y)}\:\nu(k,Y),
\end{equation}
where $\kappa$ is an undetermined constant coming from the local-noise approximation and $\nu(k,Y)$ is a Gaussian white noise satisfying
\[
\avg{\nu(k,Y)} = 0\qquad\text{ and }\quad
\avg{\nu(k,Y)\nu(k',Y')} =
\frac{2}{\abar\pi}\delta\left(\log(k^2/k'^2)\right)\delta(Y-Y').
\]

Again, most of our analytical understanding of the QCD problem comes from the fact that \eqref{eq:langevin} lies in the same universality class as the stochastic F-KPP equation. The analytical study of that equation in the weak-noise limit as well as numerical simulations \cite{simul} of \eqref{eq:langevin} show that the effect of the fluctuations is double: first, the event-by-event amplitude still displays geometric scaling but with an exponent of the saturation scale smaller than $v_c$. Then, if one considers a whole set of events, one observes a dispersion of their position in $\log(k^2)$. This means that the logarithm of the saturation scale fluctuates from one event to another, with a dispersion increasing as $\sqrt{Y}$, as expected from a random-walk process. A direct consequence of this dispersion is that, when we average over all realisations over the noise to get the physical amplitude, it leads to violations of geometric scaling.

To gain more insight on the physical consequences of these results, it is sufficient to consider an event-by-event amplitude ($r_0$ is a reference scale)
\[
T_{\text{ev}}(r,Y) = \begin{cases} 
1                  & \rho\le\rho_s\\
e^{-(\rho-\rho_s)} & \rho > \rho_s
\end{cases},\qquad 
\rho  =\log\left(\frac{r_0^2}{r^2}\right)\;\text{ and }\;
\rho_s=\log(r_0^2 Q_s^2),
\]
which satisfies unitarity constraints, colour transparency and geometric scaling, together with a Gaussian probability distribution $P(\rho_s)$ of average $\avg{\rho_s}=\log(r_0^2\bar Q_s^2)$ and dispersion $\sigma\propto\sqrt{Y}$. 

The average amplitude is then computed through
\begin{equation}\label{eq:average}
\avg{T(\rho)} = \int_{-\infty}^\infty d\rho_s\,P(\rho_s)\,T_{\text{ev}}(\rho_s-\avg{\rho_s}).
\end{equation}
At intermediate energies ($\sigma < 1$), the dispersion of the events is negligible and $\avg{T((\rho)}\approx T_{\text{ev}}(\rho_s-\avg{\rho_s})$ satisfies the geometric scaling property. However, at high energies such that $\sigma\gg 1$, dispersion starts to dominate and geometric scaling gets violated. A direct computation of \eqref{eq:average} shows that, at high energy, one has
\[
\avg{T(\rho)} = T\left(\frac{\log[r^2\bar Q_s^2(Y)]}{\sqrt{Y}}\right) = \frac{1}{2}\erfc\left(\frac{\rho-\avg{\rho_s}}{\sqrt{2}\sigma}\right)\qquad \text{for }\rho-\avg{\rho_s}\ll\sigma^2.
\]
Thus, within a window increasing linearly with rapidity, it emerges \cite{himst} a new scaling, called {\em diffusive scaling}.
Within this window, event dispersion becomes larger than the typical decrease length of $T_{\text{ev}}$. We can thus consider that, event-by-event, the amplitude is simply a Heaviside theta function $\Theta(\rho_s-\rho)$. In other words, at very high energies, the dominant contributions to the physical amplitude fully come from fronts which are at saturation {\em i.e. by black spots}. As an additional consequence, we have $\avg{T^2}=\avg{T}$ at high energy while the mean field approxiamtion predicts $\avg{T^2}=\avg{T}^2$. Note that all those results can be derived analytically \cite{strong} by assuming a strong noise in the equation \eqref{eq:langevin}.

Finally, we have to quote that the diffusive scaling property extends (as $Y$) above the average saturation momentum. The fact that we are sensitive to saturation, even when the physical amplitude is much smaller than one, can have important consequences for LHC physics {\em e.g.} for forward jet production \cite{ims}.

\end{document}